\documentclass[twocolumn,showpacs,preprintnumbers,amsmath,amssymb,eqsecnum]{revtex4}
\usepackage{graphicx}
\begin{document}
\preprint{IMAFF-RCA-05-08}
\title{New "Bigs" in cosmology}

\author{Artyom V. Yurov$^{1}$, Prado Mart\'{\i}n Moruno$^{2}$ and Pedro F. Gonz\'{a}lez-D\'{\i}az$^{2}$}
\affiliation{$^{1}$I. Kant Russian State University, Theoretical
Physics Department 14, Al.Nevsky St., Kaliningrad 236041, Russia}
\affiliation{$^{2}$Colina de los Chopos, Centro de F\'{\i}sica
``Miguel A.
Catal\'{a}n'', Instituto de Matem\'{a}ticas y F\'{\i}sica Fundamental,\\
Consejo Superior de Investigaciones Cient\'{\i}ficas, Serrano 121,
28006 Madrid (SPAIN).}
\date{\today}
\begin{abstract}
This paper contains a detailed discussion on new cosmic solutions
describing the early and late evolution of a universe that is
filled with a kind of dark energy that may or may not satisfy the
energy conditions. The main distinctive property of the resulting
space-times is that they make to appear twice the single singular
events predicted by the corresponding quintessential (phantom)
models in a manner which can be made symmetric with respect to the
origin of cosmic time. Thus, big bang and big rip singularity are
shown to take place twice, one on the positive branch of time and
the other on the negative one. We have also considered dark energy
and phantom energy accretion onto black holes and wormholes in the
context of these new cosmic solutions. It is seen that the
space-times of these holes would then undergo swelling processes
leading to big trip and big hole events taking place on distinct
epochs along the evolution of the universe. In this way, the
possibility is considered that the past and future be connected in
a non-paradoxical manner in the universes described by means of
the new symmetric solutions.
\end{abstract}

\pacs{98.80.Cq, 04.70.-s}

\maketitle

\section{Introduction}
Current research in cosmology is a very exciting, rapidly evolving
endeavor. It is actually one which however typically attracts the
kind of rejections that some scientists use to make against every
new idea or development in cosmology and that have successively
dismissed big bang, black holes, inflation, cosmic strings,
wormholes and, ultimately, accelerating expansion and phantom
energy. However, observational data are rather obstinate in favour
of the two latter suggestions and thereby of the previous one
\cite{Hannestad}. In fact, sometimes the point is not so much if a
particular theory or interpretation is too much unusual or crazy,
but, as it is believed Pauli told Heisenberg once, whether it is
unusual or crazy enough. It is in this spirit that we have
considered the subject of this paper which will deal with the
three new "bigs" that have recently arisen in the new standard
cosmological scenario: the big rip \cite{Caldwell}, the big trip
\cite{Gonz1} and the big hole. Together with the big bang and big
crunch, these new "bigs" make up the most dramatic events that may
have occurred or may eventually take place in the universe.

It appears already certain that the current universe is undergoing
a period of accelerating expansion \cite{Perlm}. What is still
under discussion indeed is whether such an acceleration is or not
beyond the barrier implied by a cosmological constant; that is
whether the parameter $w$ of the equation of state $p=w\rho$ is or
not exactly equal to -1. Actually, the case which each time
appears to be most favoured by observations is that for a value of
$w$ quite close to, but still less than -1 \cite{Hannestad}.
Nevertheless, no matter how close to -1 it could be, if $w<-1$
then our universe would be filled with some kind of phantom energy
and its expansion would be super-accelerated. Phantom energy is a
rather weird stuff indeed \cite{phantom}. Besides its nice
properties - which would include super-accelerated expansion and
the possibility to describe primordial inflation \cite{Gonz2}- it
is known to possess the following unusual characteristics. If dark
energy is described by a scalar field $\phi$ with the FRW
customary definitions, $\rho=\dot{\phi}^2/2+V(\phi)$,
$p=\dot{\phi}^2/2+V(\phi)$, with $\rho$ and $p$ the energy density
and pressure, respectively, and $V(\phi)$ the field potential,
then (i) the kinetic term $\dot{\phi}^2/2<0$ and therefore phantom
cosmologies suffer from violent instabilities and classical
inconsistencies, (ii) the energy density is an increasing function
of time which would make the quantum-gravity regime to appear also
at late times, (iii) the dominant energy condition is violated so
that $\rho+p<0$, and (iv) there will be a singularity in the
finite future dubbed big rip at which the universe ceases to exist
and near of which there may appear cosmic violations of causality.
These properties define phantom energy in the quintessence
scenario.

The present paper aims at considering big rips and big trips, so
as the newest possible phenomenon that might also occur in the
future and that we shall here denote by big hole, in the context
of new cosmological solutions where such events will all take
place not just in the future but also in the past and the
definitions of dark and phantom energy are generalized. We shall
first review in a few more detail what is understood by big rip,
big trip and big hole in usual quintessence when the equation of
state is $p=w\rho$. Using the general expression
\begin{equation}
\frac{\dot{\rho}}{\rho}=-3H(1+w)=\frac{2\dot{H}}{H},
\end{equation}
where $H=\dot{a}/a$ and $w$ is taken to be constant, we can derive
expressions for the scale factor $a(t)$ and the energy density in
the quintessence model to be
\begin{equation}
a(t)=\left[a_0^{3(1+w)/2}+\frac{3}{2}C(1+w)(t-t_0)\right]^{2/[3(1+w)]},
\end{equation}
in which $C$ is a constant and the universe speed-up anyway, and
\begin{equation}
\rho(t)= H^2
=\frac{C^2}{\left[a_0^{3(1+w)/2}+\frac{3}{2}C(1+w)(t-t_0)\right]^2}
...
\end{equation}
Now, in the phantom regime defined by $w<-1$, it can be seen that
$\rho$ increases with $t$ up to the time
\begin{equation}
t=t_* =t_0+\frac{2a_0^{-3(|w|-1)/2}}{3C(|w|-1)} ,
\end{equation}
at which it diverges. It can be seen that the scale factor $a(t)$
also blows up at $t=t_*$. This curvature singularity is what we
call big rip \cite{Caldwell}. If no pathological space-time
branches would connect the super-accelerating expanding region
before the big rip to the contracting region after it transmitting
physical information, then the big rip would mark the end of the
universe and everything in it. In the quintessence model phantom
energy will therefore be characterized by $\rho>0$, $\rho+p<0$,
$\dot{\phi}^2<0$, a big rip singularity at $t=t_*$ and a positive
definite potential
\begin{equation}
V(\phi)=\frac{1}{2}(|w|+1)C^2 e^{-3i\sqrt{|w|-1}(\phi-\phi_0)},
\end{equation}
with
\begin{equation}
\phi(t)=\phi_0
-\frac{2i}{3\sqrt{|w|-1}}\log\left[\left(\frac{a_0}{a}\right)^{3(|w|-1)/2}\right],
\end{equation}
in which $\phi_0$ is another constant.

On the other hand, it has been shown that phantom energy makes the
kind of stuff that leads to formation of Lorentzian Morris-Thorne
wormholes \cite{Lobo}. Therefore such wormholes would be expected
to copiously crop up in a universe dominated by phantom energy.
Now, once they are formed with a size small enough as to be stable
against vacuum polarization, these wormholes would start
accreating phantom energy to induce a swelling process in the
wormhole spacetime that inflates their throat so quickly that,
relative to an asymptotic observer, they would engulf the universe
itself at a time in the future given by \cite{Gonz1}
\begin{equation}
\tilde{t}= t_0 +\frac{1}{\beta+\frac{3}{2}Ab_0(|w|-1)} ,
\end{equation}
where $b_0$ is the initial radius of the spherical wormhole
throat, $\beta$ is a constant and $A$ is a numerical coefficient
of order unity. Thus, the universe would find itself passing
through the tunneling from one mouth to the other at a time that
precedes the big rip for a presumably short while and that,
depending on the relative kinematic characteristics of the
resulting insertions of the two wormhole mouths, might make the
universe travel in time toward the past or future. This traveling
is what has been dubbed big trip \cite{Gonz1} and, far from being
a catastrophic event, it really could offer a possibility of
escaping from the doomsday implied by the big rip.

Finally, let us consider what we mean here by a big hole. It is
now a matter of common wisdom that our universe contains many
black holes whose sizes ranges from a few solar masses to several
billions of solar masses. The latter black holes are called
supermassive and are thought to be placed at the galactic centres.
Any of such black holes would accrete dark energy with $w>-1$ in a
process that parallels phantom energy accretion by wormholes and
that would in principle lead to a swelling of their event horizon
so gigantic that eventually the black hole would finally swallow
the whole universe in the finite future, at a time \cite{Prado}
\begin{equation}
t_{bh}=t_0+\frac{1}{[\frac{3}{2}B \rho_0
M_0-(6\pi\rho_0)^{1/2}](1+w)},
\end{equation}
where $M_0$ is the initial mass of the black hole, $\rho_0$ is the
initial energy density and $B$ is a numerical constant of order
unity. When dealing with quintessence models, this black hole
swelling becomes an irrelevant process however. In fact, putting
reasonable astronomical data in Eq. (1.8) allows us to discover
that in this case the accretion of dark energy could only
significantly modify the black hole size if the initial black hole
mass would be of the order of the total mass of the universe, a
situation which can never be expected to occur along the entire
evolution of the universe \cite{Prado}. It will be seen
nevertheless that in the context of the cosmological models that
will be considered in this paper black holes can perfectly undergo
a big hole process.

The paper is organized as follows. In Section II we shall deal
with two new phantom cosmological solutions to represent the
space-time of an accelerating universe. What is new about such
solutions is that, in addition to showing a big rip in the future,
they also predict a big rip in the past, and that the nature of
its phantom energy is dual to that for phantom energy in usual
quintessential solution. Sec. III contains a study of the
accretion of phantom energy by wormholes and black holes in the
space-times described by the new solutions. It will be shown that
in such universes there will be multiple big trips and big holes,
depending on the cosmic time on which the observer is placed. The
possibility that the future and past be connected by means of
black hole and wormhole swellings in this kind of cosmologies is
discussed in Sec. IV. Finally, we conclude and add some comments
in Sec. V. Throughout the paper we use units such that $8\pi
G/3=c=1$.

\section{Cosmic solutions with two big rips}
In this section we shall consider two new cosmological solutions
that correspond to super-accelerated versions beyond the De Sitter
space as given in the form $a=H_0^{-1}\cosh(H_0 t)$, i.e. to
space-times that evolves both in positive and negative time and
can be shown to be symmetric relative to $t=0$.

Let us consider first a solution described by a scale factor of the
form
\begin{equation}
a(t)=\alpha\left(\beta+x\tan x\right) ,
\end{equation}
where $x=\kappa t$, $\kappa$, $\alpha$ and $\beta$ are all
positive constants. We will study this solution in the interval
$-t_R <t < t_R$, where $t_R=\pi/2\kappa$. We shall hereafter refer
to the scale factor (2.1) as solution I. In Fig 1 we give a plot
of such a solution which can be obtained by using the Darboux
transormation \cite{x1}.

\begin{figure}
\includegraphics[width=.9\columnwidth]{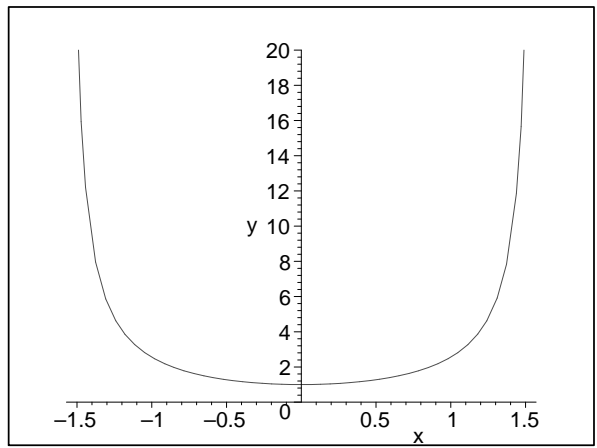}
\caption{\label{fig:epsart} Evolution of solution I with time from
a big rip in the past to a big rip in the future.}
\end{figure}

It is easy to see that $a(\pm t_R)=+\infty$, i.e. there is big rip
singularities at $t=\pm t_R$. Moreover, on the subinterval $-t_R <
t < 0$ the universe would go through a contracting stage which
ends up at $t=0$, from which it starts expanding to finally end
again up in a big rip at $t=+t_R$. Such a behavior is highly
unusual indeed. Notice as well that $a(t)$ does not vanish at any
points along the considered interval; the minimum value of the
scale factor occurs at $a(0)=\alpha\beta$. Now, we show that the
number of free parameters entering the scale factor (2.1) is
efficiently sufficient to describe our observed universe; that is
they can satisfy the following three conditions: they should
reproduce (1) the current value of the Hubble constant,
$H(t_0)=H_0=h\times 0.324\times 10^{-19}$ s$^{-1}$, where
$45<h<75$, (2) $\ddot{a}_0/a_0=7H_0^2/10$, and (3) $a_0\sim
10^{28}$ cm. We shall check in what follows that this is actually
the case. In fact, replacing the solution (2.1) into the Friedmann
equations for a flat universe, we get for the energy density
\begin{equation}
\rho=\frac{\kappa^2(T+x+xT^2)^2}{(\beta+xT)^2}
\end{equation}
and for pressure
\begin{widetext}
\begin{equation}
p=-\frac{5\kappa^2 x^2 T^4
+2x(2\beta+3)T(T^2+1)+(4\beta+6x^2+1)T^2 +4\beta+x^2}{3
(\beta+xT)^2} ,
\end{equation}
\end{widetext}
where $T=\tan x$. We can then obtain an expression for the Hubble
constant
\begin{equation}
H=\frac{\kappa(2x+\sin(2x))}{2\cos x(\beta\cos x+x\sin x)}
\end{equation}
and
\begin{equation}
\rho+3p=-\frac{4\kappa^2(1+xT)}{(\beta+xT)\cos^2 x} .
\end{equation}
Since the combination given by Eq. (2.5) is definite negative we
deduce that the strong energy condition is permanently violated and
therefore the universe undergoes constant acceleration.

Let us now consider condition (2). We can write for the constant
$\beta$
\begin{equation}
\beta=\frac{(80\cos^2 x_0 -52)x_0^2
-12x_0\sin(2x_0)+7\sin^2(2x_0)}{80\cos x_0(x_0\sin x_0+\cos x_0)} .
\end{equation}
It can be seen that at small values of $x_0$ $, \beta$ behaves
like an inverted Higgs potential (Fig. 2). The values which are of
interest here are those that lead to positive $\beta$. Such values
lie in the range $x_0 < x_m\sim 0.6$. We can then evaluate the
constant $\kappa$ to be
\begin{equation}
\kappa=\frac{7H_0\cos x_0(2x_0 +\sin(2x_0))}{40(x_0\sin x_0+\cos
x_0)},
\end{equation}
which in turn implies that
\begin{equation}
t_0=\frac{x_0}{\kappa}=\frac{40x_0(x_0\sin x_0+\cos x_0)}{7H_0\cos
x_0(2x_0+\sin(2x_0))}
\end{equation}
and $t_R=\pi/(2\kappa)$. We can now calculate the length of time
\begin{equation}
\Delta t_R = t_R-t_0 = \frac{(\pi-2x_0)t_0}{2x_0}.
\end{equation}

\begin{figure}
\includegraphics[width=.9\columnwidth]{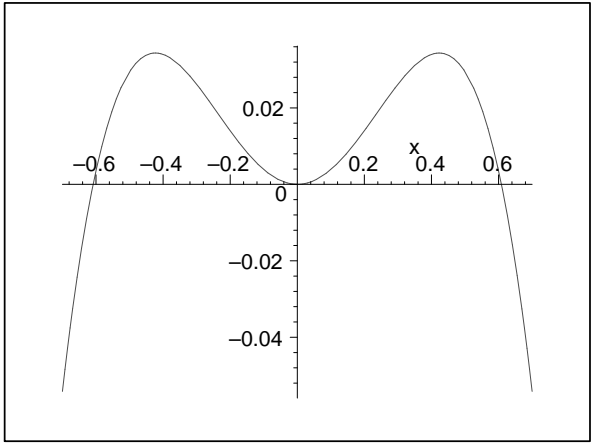}
\caption{\label{fig:epsart} Evolution of parameter $\beta$ with
time. At small $x_0$ it adopts the shape of an inverted Higgs
potential.}
\end{figure}

The results of the numerical computations are summarized in the
table. In the first row we give the different values of $x_0$ from
$0.1$ up to $0.6$; in the second row- the least values of the
scale factor, reached at $t=0$ during the change of regimes of
expanding. Next two rows present the value $t_0$ for two limiting
values of the Hubble constant: $h=45$ and $h=75$, correspondingly.
This value is related to the time elapsed since the beginning of
the universe expansion up to the present. The values of $\Delta
t_{_R}$ for $h=45$ and $h=75$ are presented in the fifth and six
rows. The last two rows contain the values computed for the age of
the universe with respect to $h=45$ and $h=75$, respectively, i.e.
the time passed from $t=-t_{_R}$ to $t=t_0$. All times in the
Table are expressed in ordinary years.

\begin{widetext}

\begin{table}
    \begin{tabular}{|c|c|c|c|c|c|c|c}\hline $x_0$ & 0.1&0.2&0.3 &0.4 &0.5& 0.6 \\\hline
 $a_{min}$&$0.27a_0$&$0.26a_0$ &$0.22a_0$
&$0.17a_0$&$0.1a_0$&$0.01a_0$ \\\hline $t_0/10^9$ (yr), h=45
&$31.4$&$32.7$&$35$&$38.3$&$43$&$49.3$
\\\hline
$t_0/10^9$ (yr), h=75 &$18.9$&$19.7$&$20.1$&$23$&$25.8$&$29.6$
\\\hline
$\Delta t_{_R}/10^9$ (yr), h=45 &$463$&$225$&$148$&$112$&$92$&$80$
\\\hline
$\Delta t_{_R}/10^9$ (yr), h=75 &$278$&$135$&$89$&$67$&$55$&$48$
\\\hline
$ T/10^9$ (yr), h=45 &$526$&$244$&$218$&$189$&$178$&$178$
\\\hline
$ T/10^9$ (yr), h=75 &$316$&$170$&$130$&$113$&$107$&$107$
\\\hline
\end{tabular}
\end{table}

\end{widetext}

A quite clearer physical motivation can be ascribed to a solution
(which will be denoted as solution II) stemming from the
Randall-Sundrum brane world model I \cite{RS} where the Fiedmann
equations can be written as
\begin{equation}
\frac{{\dot a}^2} {a^2}=\rho\left(1+\frac{\rho}{2\lambda}\right),
\end{equation}
\begin{equation}
-2\frac{\ddot
a}{a}=\rho+3p+\frac{\rho}{\lambda}\left(2\rho+3p\right). \label{2}
\end{equation}

The solution of the equations (2.10), (2.11) has the form
\begin{equation}
a^{6\epsilon}=\frac{s}{1-18\lambda\epsilon^2\left(t-t_1\right)^2},
\label{33}
\end{equation}
where
$$
\frac{p}{\rho}=w=-1-2\epsilon,
$$
$\epsilon>0$, $t_1=$const and $s=$const.

Thus, to obtain the big rip singularities one need to choose
$\lambda>0$. On the other hand, $a(t)$ must be positive so that
$s>0$. We have two big rips at
\begin{equation}
t_{\pm}=t_1\pm\frac{1}{3\epsilon\sqrt{2\lambda}}. \label{tbr}
\end{equation}
A fully symmetric solution with the big rips symmetrically displayed
around $t=0$ can immediately be obtained by choosing $t_1=0$

The energy density and pressure will be:
\begin{equation}
\rho=-\frac{2\lambda}{1-18\lambda\epsilon^2\left(t-t_1\right)^2}<0,
\label{rho}
\end{equation}
\begin{equation}
p=\frac{2\lambda(1+2\epsilon)}{1-18\lambda\epsilon^2\left(t-t_1\right)^2}>0.
\label{p}
\end{equation}
Now, let consider the universe filled with scalar field $\phi$, such
that
$$
\rho=K+V,\qquad p=K-V,
$$
with $K={\dot\phi}^2/2$, $V=V(\phi)$. Using (\ref{rho}) and
(\ref{p}) one gets
\begin{equation}
K=\frac{2\lambda\epsilon}{1-18\lambda\epsilon^2\left(t-t_1\right)^2}>0,
\label{K}
\end{equation}
and
\begin{equation}
V=-\frac{2\lambda(\epsilon+1)}{1-18\lambda\epsilon^2\left(t-t_1\right)^2}<0.
\label{V}
\end{equation}
Thus, in the brane world we have the highly nontrivial situation
that a model with negative potential (and positive $K$) results in
a big rip singularity.

One can see that
$$
p+\rho=\frac{4\lambda\epsilon}{1-18\lambda\epsilon^2\left(t-t_1\right)^2}>0,
$$
and
$$
3p+\rho=\frac{4\lambda(1+3\epsilon)}{1-18\lambda\epsilon^2\left(t-t_1\right)^2}>0.
$$
But since $\rho<0$ both weak and strong energy conditions are
violated. It is interesting to note that
$$
\frac{\ddot
a}{a}=\frac{6\epsilon\lambda\left(6\lambda\epsilon(t-t_1)^2+18\lambda\epsilon^2(t-t_1)^2+1\right)}
{1-18\lambda\epsilon^2\left(t-t_1\right)^2}>0.
$$
We restore finally $V$ as a function $V(\phi)$. To do this one
need to find $\phi=\phi(t)$ from Eq. (2.16), expressing then
$t=t(\phi)$ to finally substitute it into Eq. (2.17). After simple
calculation one get
$$
\phi(t)=\phi_0\pm \frac{1}{3}\sqrt{\frac{2}{\epsilon}}{\rm
arcsin}\left(3\epsilon\sqrt{2\lambda}(t-t_1)\right),
$$
and hence
\begin{equation}
V(\phi)=-\frac{2\lambda(\epsilon+1)}{{\rm
cos}^2\left(3\sqrt{\frac{\epsilon}{2}}(\phi-\phi_0)\right)},
\label{VV}
\end{equation}
where $\phi_0=$const. Note that in this case the field $\phi$ is
real in spite of corresponding to a phantom stuff.

Thus, we have a model with positive tension, positive $K$, negative
potential $V$, negative $\rho=K+V$ and positive pressure $p=K-V$. At
the same time, $\rho+p>0$ ($\rho+3p>0$ too) and ${\ddot a}/a>0$.
This universe is filled with a scalar field with the rather amusing
potential (\ref{VV}).

We have two big rips at $t=t_{\pm}$ (see Eq. (\ref{tbr})). In the
''classical'' limit $\lambda\to\infty$ we have only one big rip at
$t=t_1$. Therefore, the situation with two big rips (initial and
final) should be taken to be a brane effect.

It has been seen that inserting a parameter $w<-1$ in this brane
world model leads to a distinct phantom stuff, one which is dual to
the phantom energy resulting in quintessence with $w<-1$ in that in
the present case it is the energy density $\rho$ but not $\rho+p$
what is negative definite. Moreover, dual phantom is associated with
a scalar field with negative potential and positive kinetic term.
The only prediction which is shared by phantom and dual phantom is
the emergence of future (or past) singularities.

Actually usual phantom energy with negative scalar-field kinetic
term appears to occur only in the regions before the first big rip
at $t_-$ and after the second big rip at $t_+$ (see Fig. 3). Had
we taken $w>-1$ in the precedent calculation then the resulting
scale factor had described a universe initially contracting down
to a singularity (a big crunch respect to an interior observer or
a big bang respect to the exterior, following region) at $a=0$,
both for positive and negative time, followed in the two branches
by usual accelerating regions which extend toward infinity. It is
worth noticing that in this case, whereas the exterior
accelerating regions are filled with conventional dark energy, in
the region between the two zeros dark energy would be also unusual
in that it would be the dual (with negative values for $\rho$ and
$\rho+p$) to conventional dark energy (Fig. 3). All the regions
with dual stuffs would disappear in the limit
$\lambda\rightarrow\infty$ where no brane is present. Therefore
the existence of such regions and hence of a second big rip can be
considered to be a pure brane effect.

\begin{figure}
\includegraphics[width=.9\columnwidth]{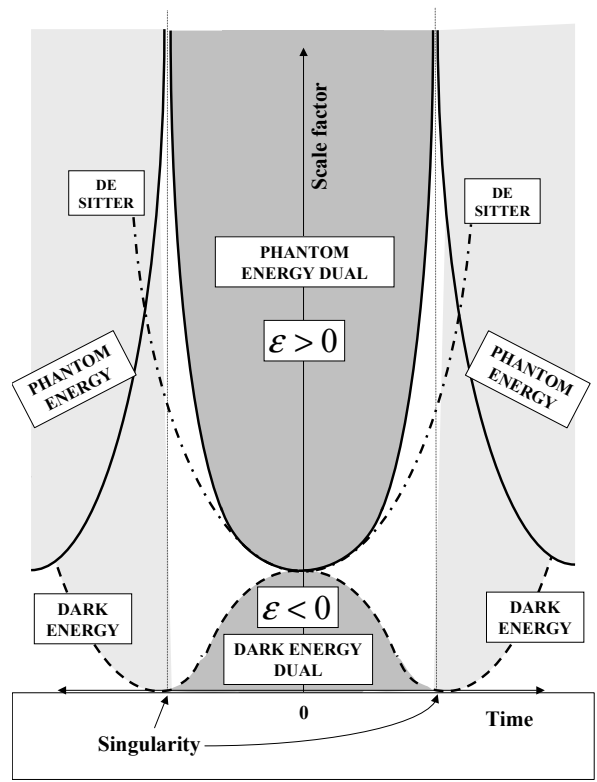}
\caption{\label{fig:epsart} The different dark-energy and
phantom-energy regions which can be obtained for Solution II. Dual
regions correspond to the interval between the two symmetric (big
bang or big rip) singularities and are originated from pure brane
effects.}
\end{figure}

We note finally that at first sight it could seem unnatural to
talk about big rip singularities and accelerating universe in a
model of brane world which would be expected to describe the early
universe. However, since the absolute value of the energy density
is an increasing function of squared time in the present model,
the brane and quantum effects are here expected to become relevant
at late time, instead of at the early stages.

The brane symmetric solution, on the other hand, has a rather
surprising property which can help to create and maintain a
shortcut for interstellar travel. In fact, following Krasnikov
\cite{Kras1} one can define a shortcut as follows. Let $C$ be the
timelike cylinder in Minkowski space $L^4$, $M$ a globally
hyperbolic spacetime and $U$ a subset of $M$. Then $U$ will be a
shortcut if the isometry $\kappa: (M-U)\to(L^4-C)$ and two points
$p$ and $q$ exist such that $p\preceq q$ (i.e. there is a
future-directed timelike curve from $p$ to $q$) and this is not
the case for $\kappa(p)$ and $\kappa(q)$. Examples of shortcuts
are wormholes (like the one we are going to consider in the next
section) and the so-called Krasnikov tube \cite{Krastube},
\cite{Krastube2}. In all the cases where a shortcut takes place
the weak energy condition (WEC) $T_{\mu\nu}t^{\mu}t^{\nu}\ge 0$
(where $T_{\mu\nu} $ is the stress-energy tensor and $t^{\mu}$ is
any timelike vector) must be violated. Now, even though violation
of WEC in quantum field theory can be an allowable phenomenon
(like in Casimir effect) the {\it artificial creation} of
shortcuts (for example, by future advanced civilizations) is
rather problematic because of the following reason. Ford and Roman
(\cite{11}, \cite{12}; see also \cite{13}, \cite{14}) showed that
in the case of $d=2$ massless scalar fields the following
inequality holds
\begin{equation}
\rho_f\equiv \int_{\tau_1}^{\tau_2}
d\tau\rho(\tau)f(\tau-\tau_0)\ge -|\tau_2-\tau_1|^{-d},
\label{ineq}
\end{equation}
where $\rho=T_{{\hat 0}{\hat 0}}$ (''hats'' mean that one use the
orthonormal basis), $d$ is dimension of the spacetime and $f$ is a
function such that $f\in C^{\infty}$, ${\rm supp}
f\in(\tau_2,\tau_1)$ and~\footnote{The additional condition
$|\tau_2-\tau_2|\le \left({\rm max}|R_{{\hat
\mu}{\hat\nu}{\hat\rho}{\hat\sigma}}|\right)^{-1/2}$ must also
hold.}
$$
\int _{\tau_1}^{\tau_2} d\tau \frac{(f'(\tau))^2}{f(\tau)}\le 1.
$$

It is currently believed that this inequality will apply for all
cosmic solutions and shortcuts. In Ref. \cite{Kras1} (see also
\cite{Krastube2}, \cite{15}) it was actually shown that if the
inequality (\ref{ineq}) holds then one would need to have
$E/c^2=3\times 10^{62}$ kilograms of negative energy to construct
a shortcut able to allow for the translation of an object with a
size about 1 meter. Thus the condition (\ref{ineq}) demonstrates
that future manufacturers of shortcuts will meet with serious,
rather unsurmontable difficulties. However, if we consider
solution (2.12) and take $t_*$ as the absolute value of the time
at which future big rip takes place, then using Eq. (2.14) one
gets
$$
\lim_{_{\tau_2\to t_*}}\int_{\tau_1}^{\tau_2}
d\tau\rho(\tau)f(\tau-\tau_0)=-\infty .
$$
Therefore inequality (\ref{ineq}) no longer applies in the case
that we add phantom energy to a RS brane type I, so that the above
alluded difficulties for constructing shortcuts in the future are
largely relaxed.

\section{Big trips and big holes in the symmetric solutions}
Apart from their intrinsic interest, the cosmic solutions
considered in Sec. II could allow in principle for the following
possibility. It is known that as one goes toward the big rip
singularity there would occur the process of a fast wormhole
swelling taking the size of the wormhole throat to infinity before
the universe reaches the big rip \cite{Gonz1}. The point now is:
does such a wormhole swelling take also place as one is
approaching the big rip at negative time? If this question is
answered affirmatively then it would be only natural to suppose
that in the distant future a space-time bridge could be formed
reversibly linking the universe about to get in the future big rip
with the same universe shortly after the past big rip. Or in other
words, we can then have a model which bears a striking similarity
with the famous G\"{o}del model allowing for closed timelike curves
\cite{Goedel}. Moreover, if instead of a wormhole swelling we
would have a black hole swelling leading to what we have dubbed as
a big hole, then the bridge between future and past could also be
formed, this time in an irreversible fashion. In what follows we
shall investigate this problem by considering the phantom energy
accretion onto wormholes and black holes in the framework of the
cosmological solutions discussed in Sec. II.

There is a simple argument which seems to preclude the existence of
a symmetric couple of wormhole swellings, both leading to big trip
in the symmetric solution I. That argument runs as follows.
Similarly to as an antiparticle is nothing but its counter particle
moving backward in time, the exotic mass of a wormhole should be
seen as just ordinary matter in a Schwarzschild wormhole evolving in
an external negative time. However, a wormhole with ordinary matter
is known to be unstable and pinches off immediately to convert
itself into a black hole plus a white hole. Thus, a wormhole with
positive mass evolving backward in time is expected to be unstable
by the following argument. As one is going backward in time the
first of the considered models becomes exactly equivalent to a
phantom model moving forward in time due to the symmetric character
of the solution, and therefore the positive energy density and
curvature tend to infinite as $t$ tends to $-t_R$. It follows that
as one is approaching $-t_R$ there will be a phantom energy flow
into the wormhole which makes its positive energy to decrease and
really vanish at the singularity at $-t_R$. This is the instability
that prohibits a big trip to take place on the negative time branch
of the solution. Whether or not one would take this instability to
be the same as that taking place in black holes in a quintessential
phantom universe becomes a matter of interpretation. This heuristic
prediction can indeed be confirmed by calculation. In fact, if we
write the Friedmann equations for a flat universe as
\begin{equation}\label{uno}
\left(\frac{\dot{a}}{a}\right)^2=\rho
\end{equation}
\begin{equation}\label{dos}
2\frac{\ddot{a}}{a}=-(\rho+3p)
\end{equation}
By rearranging and substituting in these equations, we have:
\begin{equation}\label{tres}
\frac{3}{2}(p+\rho)=-\frac{\ddot{a}}{a}+\left(\frac{\dot{a}}{a}\right)^2
\end{equation}
Now, the known expression for the rate of mass of a wormhole due
to phantom energy accretion is \cite{Gonz1}
\begin{equation}\label{cuatro}
\dot{m}=-\frac{3}{2} A m^2(p+\rho)
\end{equation}
Integrating this rate equation with the use of Eqs. (3.3) and
(3.8), we obtain
\begin{equation}\label{cinco}
m=m_0\left(1-Am_0\left(\frac{\dot{a}(t)}{a(t)}-\frac{\dot{a}(t_0)}{a(t_0)}\right)\right)^{-1}
\end{equation}

We consider next the scale factor
\begin{equation}\label{seis}
a(t)=\alpha (\beta+\kappa t\cdot tan(\kappa t))
\end{equation}
where $\alpha$, $\kappa$ and $\beta$ are all positive constants.
We will study solution (3.6) on the interval $-t_R<t<t_R$, with
$t_R=\pi/(2\kappa)$. It is easy to see that $a(\pm t_R)=+\infty$,
so that the Universe starts at a big rip singularity and
thereafter goes along a stage of contraction on $-t_R<0<t$ which
ends at $t=0$; then it starts expanding, finally ending again at a
big rip singularity. With this scale factor we can obtain:
\begin{equation}\label{siete}
m(x)=m_0\left[1+Am_0H_0-Am_0\kappa\frac{2x+sin(2x)}{\beta
cos²x+xsin(2x)}\right]^{-1} ,
\end{equation}
where we have defined $x=\kappa t$, $-\pi/2<x<\pi/2$. Then
$m(x\rightarrow \pm \pi/2)=0$. It is trivial to notice that
$m(x)\neq m(-x)$, i. e., the growth of the wormhole does not
preserve the symmetric character of the scale factor under time
reversal. As we shall see now this asymmetry leads to the
emergence of a big trip (i.e. a blow-up of the wormhole throat
size) only on the positive-time branch, not on the negative-time
branch. In order to study the emergence of possible divergences in
the case of solution I, we subdivide the interval in three pieces:

\paragraph{Subinterval $-\pi/2<x<0$} Here we can express the mass as
$$\frac{m(x)}{m_0}=\left[1+Am_0(F(x)+H_0)\right]^{-1} ,$$
where
$$F(x):=-\kappa\frac{2x+sin(2x)}{\beta cos²x+xsin(2x)}>0 .$$
If there is a wormhole within the Universe, then it will always have
a finite size on that subinterval.

\paragraph{Point x=0} At this point the wormhole throat will have a
finite size which equals
$m(0)=m_0(1+Am_0H_0)^{-1}$.

\paragraph{Subinterval $0<x<\pi/2$} In this case, we have
$$\frac{m(x)}{m_0}=\left[1+Am_0H_0-Am_0G(x)\right]^{-1}  ,$$
where $$G(x):=-F(x)>0 .$$ It follows that if the wormhole grows
infinitely somewhere, it will be along this subinterval. The
divergence points will be the zeros of de function $J(x)$, with
\begin{widetext}
$$J(x)=(1+Am_0H_0)\beta cos²x+[(1+Am_0H_0)x-Am_0\kappa]sin(2x)-Am_0\kappa 2x  .$$
\end{widetext}
We can see that, since $J(0)>0$, $J(\pi/2)<0$ and $J(x)$ is
continuous; then there necessarily is at least one zero on this
subinterval (if there would be more than one zero, then they must
be a odd number; but we think that, because of the regular
recurrence of the function, there are not more of one zero on
$0<x<\pi/2$). That is, there is a big trip here but, as we have
seen, there cannot be another one symmetric to this on the
subinterval defined for negative time.

We note however that, even though there is no big trip on negative
time, a big hole should be displayed on negative time in spite of
the feature that we are dealing with a dark stuff with $w<-1$,
where no big hole would be in principle expected. If we have a
black hole of mass $M$, then the rate of change of $M$ due to the
accretion is given by \cite{Babichev}
\begin{equation}\label{cuatro}
\dot{M}=\frac{3}{2} B M^2(p+\rho) ,
\end{equation}
with $B$ another numerical constant whose value is of the order $A$.
In this case we obtain for the black hole mass
\begin{equation}\label{siete}
M(x)=M_0\left[1-Bm_0H_0+BM_0\kappa\frac{2x+\sin(2x)}{\beta
\cos²x+x\sin(2x)}\right]^{-1} .
\end{equation}
Clearly, the zero of the denominator in this expression takes now
place on the subinterval on negative time (first of the above
subintervals), but not on the subinterval on positive time. This
indicates that a big hole must take place. Unlike in quintessence
models, such a phenomenon would have quite relevant effects in
this case and proceed up to completion for any set of reasonable
parameters. In Fig. 4 we schematically show the processes of big
trip and big hole for solution I.

\begin{figure}
\includegraphics[width=.9\columnwidth]{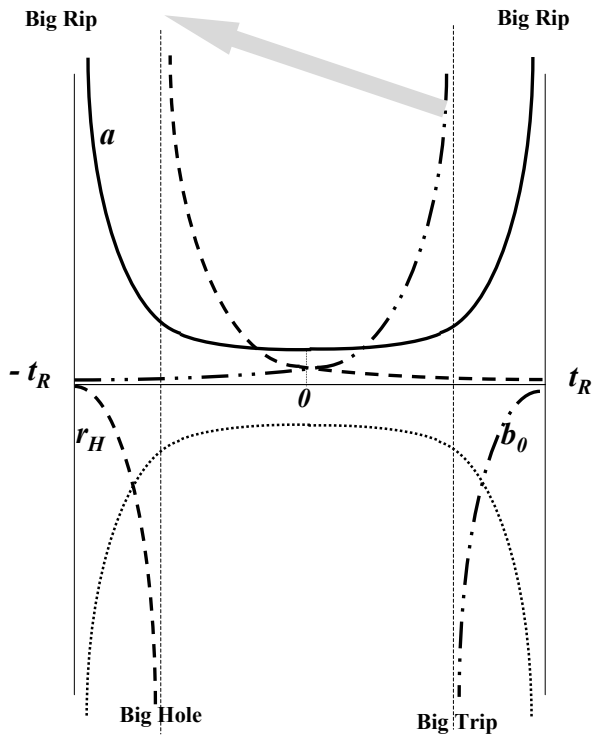}
\caption{\label{fig:epsart} Relative placements along time of the
big trip and big hole processes which take place for solution I.
It can be observed that whereas a big trip of wormhole occurs on
positive time, the big hole of black hole takes place on negative
time.}
\end{figure}

We shall consider in what follows the emergence of big trip events
in case of solution II. The rate of change of the wormhole throat
radius in the case of a brane symmetric solution ($t_1=0$) is
\begin{equation}
\dot{b}=3A\epsilon\rho=\frac{6A
\epsilon\lambda}{18\epsilon^2\lambda t^2-1},
\end{equation}
which, after integrating, yields
\begin{equation}
b=\frac{b_0}{1-\frac{A\sqrt{2\lambda}b_0}{2}\ln\left[\frac{(1-3\sqrt{2\lambda}\epsilon
t)((1+3\sqrt{2\lambda}\epsilon t_0)}{(1+3\sqrt{2\lambda}\epsilon
t)(1-3\sqrt{2\lambda}\epsilon t_0)}\right]} .
\end{equation}
The zeros of the denominator of this function would take place at
\begin{equation}
\tilde{t}=\frac{1-\xi}{1+\xi}t_* ,
\end{equation}
where $t_*=1/(3\sqrt{2\lambda}\epsilon)$ is the absolute value of
the time at which the big rips take place, and
\begin{equation}
\xi=\left(\frac{1-\frac{t_0}{t_*}}{1+\frac{t_0}{t_*}}\right)e^{2/(A\sqrt{2\lambda}b_0)}
...
\end{equation}
It follows that (1) if $0<t_0<t_*$, $\xi>1$, then there will be a
divergent wormhole swelling on $-t_* <t<0$ (a situation that also
corresponds to $0<|t_0|<t_*$, with $\xi<0$); (2) if $0<t_0<t_*$,
$0<\xi<1$, then there will be a divergent wormhole swelling on $t_*
>t>0$; (3) if $t_0>t_*$, $\xi<0$ $|\xi|>1$, then there will be a
divergent wormhole swelling on $t<-t_*$ (a situation that also
corresponds to $|t_0|>t_*$, with $\xi<-1$, $|\xi|>1$); and (4) if
$t_0>t_*$, $\xi<0$ $|\xi|<1$, then there will be a divergent
wormhole swelling on $t>t_*$. Hence, there will be a big trip on
every of the possible regions in which the whole time interval
running from $-\infty$ to $+\infty$. A similar, but not the same
situation is also attained in case that we have a black hole
instead of a wormhole. In this case, the rate of change of the
black hole mass $M$ due to accretion is
\begin{equation}
\dot{M}=- 3B\epsilon\rho=-\frac{6
B\epsilon\lambda}{18\epsilon^2\lambda t^2-1},
\end{equation}
with which the following time-dependent black hole mass can be
derived
\begin{equation}
M=\frac{M_0}{1+\frac{
B\sqrt{2\lambda}M_0}{2}\ln\left[\frac{(1-3\sqrt{2\lambda}\epsilon
t)((1+3\sqrt{2\lambda}\epsilon t_0)}{(1+3\sqrt{2\lambda}\epsilon
t)(1-3\sqrt{2\lambda}\epsilon t_0)}\right]} .
\end{equation}
Thus, $M$ will diverge (big hole) at exactly the times
\begin{equation}
\bar{t}=\frac{1-\eta}{1+\eta}t_* ,
\end{equation}
with
\begin{equation}
\eta=\left(\frac{1-\frac{t_0}{t_*}}{1+\frac{t_0}{t_*}}\right)e^{-2/(B\sqrt{2\lambda}M_0)}.
\end{equation}
Relative to the distinct observers we obtain in this way a set of
possible big holes distributed along the entire interval. The
precise pattern of such big holes, together with that for big
trips is displayed in Fig. 5. Il will be discussed in the next
section how all the regions of the complete time interval can be
connected to each other without passing through the big rip
singularities. It has been pointed out that, contrary to what is
currently observed, braneworld scenarios do not allow the
existence of static black hole \cite{Maartens}. At first sight,
this could make the above calculation on black holes irrelevant.
However, we are using a kind of nonstatic black holes which might
be allowed in braneworlds and, moreover, one can always take our
model as an approximation where any kind of black holes can be
added.

\begin{figure}
\includegraphics[width=.9\columnwidth]{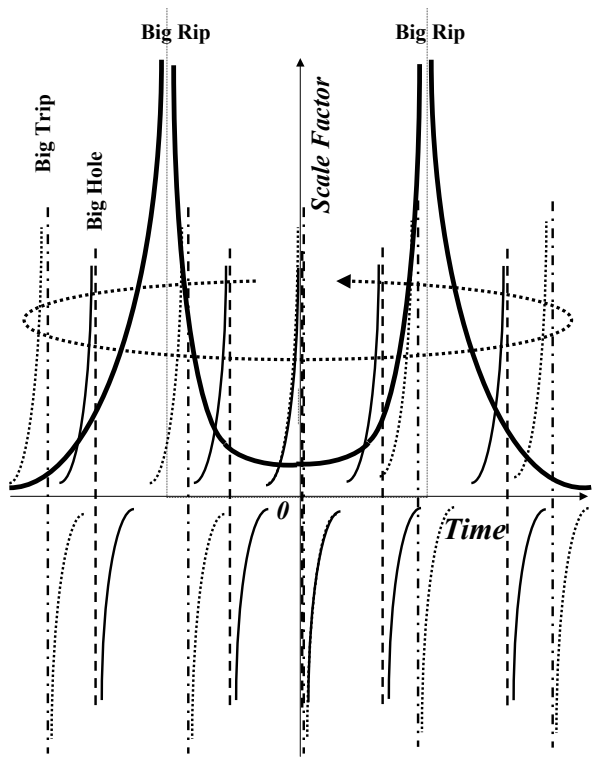}
\caption{\label{fig:epsart} Distribution of the processes of big
trip and big hole over the time for solution II. The exact
positions will depend on the initial parameters characterizing the
corresponding hole, $b_0$ for the wormholes and $M_0$ for the
black holes, and on the time at which the observer is placed.}
\end{figure}

\section{Bridges to the past and future}
A Morris-Thorne wormhole can be converted into time machine
allowing any object traversing it to time travel into the past or
future when the two wormhole mouths are provided with a given
relative motion \cite{Thorne}. Thus, since the space-time where
the mouths of such a wormhole are inserted has a given expanding
or contracting dynamics, one should expect that a swelling
wormhole might behave like a time machine and, when sufficiently
grown up, it could even make the swallowed universe and everything
in it to time travel as well. It has been in particular shown
\cite{Gonz1} that during the wormhole swelling induced by phantom
energy accretion the chronology protection conjecture
\cite{Hawking} ic violated, so that macroscopic wormholes become
quantum-mechanically stable during that accretion process.
According to the distinct processes considered in the previous
section we can have different kinds of bridges connecting the past
and future of the universe, circumventing or not a singularity
type big rip, big bang or big crunch. In order to determine the
structure and properties of these bridges, we have to take into
account two requirements: (1) Any of the space-time swelling
processes we have considered in this paper must be described for
an asymptotic observer at radial coordinate $r\rightarrow\infty$,
for otherwise such processes simply cannot take place or would
lead to quite different behaviours \cite{Gonz1}, and (2) by their
very definition, the spacetime of both a Morris-Thorne wormhole
and a black hole ought to be asymptotically flat. This condition
makes it impossible that the universe can travel along its own
time through a single wormhole. We thus distinguish the following
processes.

\noindent {\bf Single wormhole processes} While still smaller than
the host universe, the swelling wormhole may allow for given
amounts of energy to time travel. Once the wormhole has grown up
larger than the cosmological horizon it can no longer insert its
mouths into the host universe and, in order to become implanted
somewhere, it must necessarily making recourse to other different,
sufficiently larger universes, if they are at all available (for
example in a multiverse scenario (see Fig. 6)), while satisfying
the Israel junction conditions. The lack of a common time for the
assumed set of universes would convert the big trip into a simple
energy transfer process without any violation of causality. There
is thus no proper time travel in the latter case.

\begin{figure}
\includegraphics[width=.9\columnwidth]{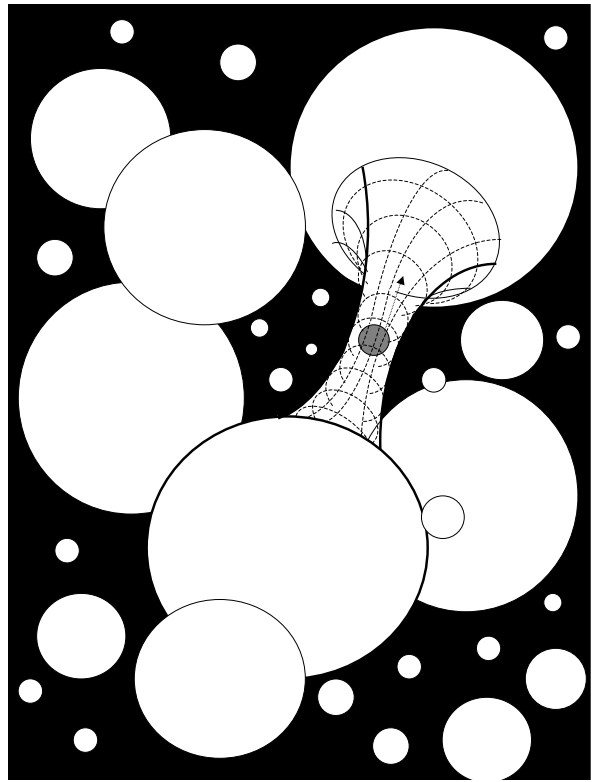}
\caption{\label{fig:epsart} Pictorial representation of the big
trip process when it is carried out by a single grown-up wormhole
within the framework of a multiverse picture. In this case the
universe does not travel along its own time but behaves like
though if its whole content were transferred from one different
larger universe to another, also larger universe.}
\end{figure}

\noindent {\bf Processes induced by a couple of swelling
wormholes} As a couple of wormholes, one in the past and the other
in the future, are growing bigger than the universe in which they
are implanted, these wormholes should cease to insert their mouths
onto the large regions of the universe where they were originally
inserted. The resulting open mouths of one of the wormholes can be
then connected to the resulting open mouths of the other, so that
the two wormholes turn out to be mutually connected to each other
in such a way that they form up a compact, closed quickly
inflating tunnel during the while when their throats have grown
beyond the size of the cosmological horizon. The universe trapped
inside can then flow along the resulting traversable closed
tunnel, travelling in this way along its own time (see Fig. 7).
This "big kiss" can always be made to satisfy the Israel junction
conditions and therefore effectively allows for the existence of
connections between the past and future of the universe. Once the
wormholes are annihilated just after the big trip by converting
themselves into a couple of black/while hole pairs, the universe
would continue its conventional causal evolution, re-starting at
the moment it has finished its time travelling.

\begin{figure}
\includegraphics[width=.9\columnwidth]{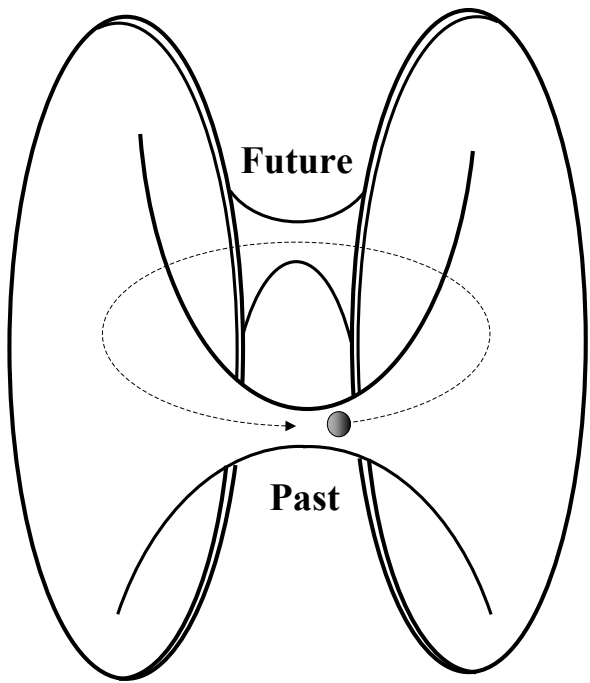}
\caption{\label{fig:epsart} Pictorial representation of the big
trip process when two grown-up wormholes are used, one in the past
and the other in the future. In this case the two wormhole connect
their mouths in such a way that they form a compact tunnel through
which the universe can travel along its own time, into the past or
future.}
\end{figure}

\noindent {\bf Processes induced by the combined action of a
swelling wormhole and a swelling black hole} The topology resulting
in this case is like that of the two swelling wormholes considered
in the precedent situation, unless by the feature that one of the
two wormholes is replaced for a swelling black hole. The effect is
the same, too: the universe can time travel along its own time, this
time by using a black hole in the past or future as an intermediate
stage on which the wormhole is inserted in a way that also satisfies
the junction conditions. At first sight, it could be expected that
the black hole would rapidly accrete all the exotic energy that
keeps the wormhole throat open, so making impossible this kind of
process, but that does not happen in any the cases considered in
Sec. III because the black hole swelling takes place only on the
branch of time (negative or positive) where its surface gravity
becomes always repulsive relative to an observer moving (forward or
backward, respectively) in time.

Thus, the question posed at the beginning of Sec. III can be
responded in the affirmative. Bridges linking the past and future
of the universe may be formed that mark the itineraries of closed
timelike curves in a way that is somehow reminiscent to the dream
that G\"{o}del kept for most of his late life. Moreover, wormholes by
themselves may serve as ways of escaping from the initial and
ripping singularities. However, the question of what happens with
causality should still be addressed in such a scenario. In fact,
the connection between past and future in our universe may lead to
time travel paradoxes (TTP); that is, you for example can kill
your parents before you were born (grandfather paradox) or you can
kill yourself in the past~\footnote{After all, this seems more
humane than killing one's parents!}. Could paradoxes like these be
avoided in our scenario? There actually are many articles dealing
with this matter but the solution is still unclear. For example,
Krasnikov \cite{Kras} has defined time travel paradoxes in
physical terms, concluding that no paradoxes arise in general
relativity. Another possible way to solve TTP is connected with
the many worlds interpretation of quantum mechanics
\cite{Deutsch}, but this solution is faced with its own problems
too \cite{Everett}.

A very attractive solution to the problem of TTP was suggested in
Ref. \cite{Novikov}. For the case of a ''hardsphere''
self-interaction potential and wormhole time machines, it was
showed that for a particle with fixed initial and final positions
traversing the wormhole just once, the only possible trajectories
minimizing the classical action are those which are globally
self-consistent. The principle of self-consistency (originally
introduced by Novikov) becomes thus a natural consequence from the
principle of minimal action. Although the verdict is still not
settled down on the general validity of this solution (see, for
example, \cite{Everett}), we shall assume it to be correct. Then
it will be shown in what follows that the flatness problem can be
solved without making recourse to the inflationary paradigm. In
order to understand how this is possible, let us consider the
one-particle Schr\"odinger equation
\begin{equation}
i\hbar\frac{\partial\Psi}{\partial
t}=-\frac{\hbar^2}{2m}\nabla^2\Psi+V\Psi. \label{Schrod}
\end{equation}
Using Bohm substitution \cite{Bohm} $\Psi=R{\rm e}^{iS/\hbar}$
(with real functions $R$ and $S$) we can obtain two equations. The
first one will be just the Hamilton-Jacobi (H-J) equation for a
single particle moving in a quantum potential
$$
U=V-\frac{\hbar^2}{2m}\frac{\nabla^2 R}{R}.
$$
The classical principle of minimal action will hold if $U=V$. In
fact, classical mechanics assumes that a particle executes just
single paths between two points basing on the minimization of the
classical action $S$. In quantum theory all possible paths
contribute to the path integral. Thus the principle of minimal
action will hold in the present case if $\nabla^2 R=0$. On the
other hand, the latter equation has no bounded nonsingular
solutions unless we have $R=$const~\footnote{Now we must extend
the Hilbert space to a Rigged Hilbert space that includes delta
functions}. The simplest example is the plane waves in
nonrelativistic quantum theory. The wave function describes then a
free particle.

We can perform a parallel development for the Wheeler-DeWitt
equation ${\hat H}\psi=0$, where ${\hat H}$ is the
super-Hamiltonian operator and $\psi$ is the wave function of the
universe. Therefore we will have $\psi=R{\rm e}^{iS/\hbar}$, where
$R=$const. In other words, $R$ does not depend on the scale factor
$a$. Such a non-normalizable wave function of the universe was
already suggested by Tipler in \cite{Tipler} by introducing the
boundary condition that the quantum state of the universe should
allow Einstein equations to hold exactly in the present epoch. One
can therefore conclude that this boundary condition must be
satisfied in order to avoid the emergence of TTP in cosmology,
provided the conclusion in \cite{Novikov} is correct. We note
however that in the neighborhood of the big rip singularities
Einstein equations cannot hold and therefore TTP could then take
place.

Moreover, by using a wave function with $R=$const Tipler also
showed how the flatness problem can be solved. To see this, let us
consider the probability that we will find ourselves in a closed
universe with radius larger than any given radius $a_{given}$
$$
\int_{a_{given}}^{+\infty}|\psi|^2da=+\infty,
$$
whereas
$$
\int_0^{a_{given}}|\psi|^2da < +\infty,
$$
so the {\it relative} probability that we will find ourselves in a
universe with radius larger than any given radius $a_{given}$ is
one. Thus, using the condition that {\it there are no TTP} one can
also solve the flatness problem in cosmology.

To make this point clearer let us return to the case of
(\ref{Schrod}). If $V=0$ then the solution of (\ref{Schrod}) has
the form:
\begin{equation}
\displaystyle{ \Psi={\rm e}^{-i\left(\hbar{\vec k}^2t/(2m)+{\vec
k}{\vec r}\right)}.} \label{free}
\end{equation}
In the lab we have prior information about initial location of the
particle so one must consider wave packet rather then plane wave
(\ref{free}). But if we do not have any (prior) information then
we are compelled to use the non-normalizable wave function
(\ref{free}). In this case we cannot calculate the probability
$p(V<V_{given})$ to find the particle inside the given volume
$V_{given}$ and one can conclude that $ p(V<V_{given}) \sim
V_{given}$. The probability to find this particle out of this
volume but inside the volume $V_1$ will be proportional to
$V_1-V_{given}$. Now a {\it relative} probability can be obtained
exactly:
$$
p_{rel}(V<V_{given})=\frac{ V_{given}}{ V_1-V_{given}}.
$$
It follows that the relative probability to find the particle
inside of the volume $V_{given}$ ($V_1\to+\infty$) will be zero
and the relative probability to find this particle outside this
volume will be one. {\it This is a direct consequence from not
having any prior information}.

This situation is unusual for the lab but usual for quantum
cosmology. In the latter case we do not have any prior information
and one need to use the non-normalizable wave function of the
universe $\psi$ instead of, say, a wave packet. It is well known
that the wave function of the universe is non-normalizable in
tree-level approximation and this problem can be solved by
including loop effects. But in the framework of our above
approach, the wave function in tree-level approximation becomes
the true wave function and all loop effects must be suppressed.
Our conclusion therefore is that when avoiding time travel
paradoxes the way we have outlined above, one finds an extra
unexpected reward: the solution of the flatness problem without
making any recourse to inflationary ideas.

\section{Conclusions and further comments}
This paper deals with new symmetric cosmological solutions for the
late or early universe by using dark energy stuffs that satisfy
the energy conditions or violate them in several different ways.
The distinctive property of all such solutions is that they can
double in a symmetric manner the main singular events predicted by
the corresponding quintessential cosmologies. In particular, they
make the big rip and the big bang (or big crunch) to appear twice,
one on the positive branch of time and the other on the negative
one. However the effects of the accretion of this generalized dark
energy onto black holes and wormholes do not display such a
symmetry so that the big trip only appears once, either on the
positive branch or on the negative branch of time. An interesting
aspect of our work resides on the possibility of making viable a
relevant swelling of black hole space-times so that it may lead to
the here dubbed big hole phenomenon by which the black hole grows
so big as to swallow the whole universe. That big hole was
precluded in quintessence models and is also predicted to appear
here just once, either on positive or negative time.

The most interesting sector of the above solutions are derived
from the brane world scenario and it is shown that the prediction
of a second singular events is a pure brane effect and that the
late evolution of the universe predicted in such brane models is
due to the existence of an energy density whose absolute value
increases with cosmological time, so making most relevant the
brane and quantum effects to appear at the latest times. On the
other hand, the simultaneous emergence of the big trip and big
hole phenomena before and after the cosmic singularities makes it
possible to circumvent such singularities so that the full
interval for the universe evolution is extended to cover the
entire range from $t=-\infty$ to $t=+\infty$.

It should also be remarked that in case that $\epsilon<0$ in the
symmetric solution derived from the brane model (solution II), the
two symmetric zeros of the scale factor at $t_{\pm}=\pm
t_{bb}=1/(3\sqrt{2\lambda}|\epsilon|)$ actually correspond to two
big bang (or big crunch) singularities where the energy density
diverges. Such singularities can be also circumvented as,
similarly to as it happens for positive $\epsilon$, wormhole and
black hole swellings and its corresponding big trip and big hole,
now taking place at
\begin{equation}
t=T=\frac{\xi-1}{\xi+1}t_{bb} ,
\end{equation}
with
\begin{equation}
\xi=\left(\frac{1+\frac{t_0}{t_{bb}}}{1-\frac{t_0}{t_{bb}}}\right)
e^{2/(A\sqrt{2\lambda}b_0)}
\end{equation}
for wormholes and
\begin{equation}
\xi=\left(\frac{1+\frac{t_0}{t_{bb}}}{1-\frac{t_0}{t_{bb}}}\right)
e^{-2/(B\sqrt{2\lambda}M_0)}
\end{equation}
for black holes, would again crop up along the entire time
interval from $-\infty$ to $+\infty$ in such a way that, relative
to different observers placed on distinct regions of that
interval, there will be no need to pass through the big bang (or
big crunch) singularities. This mechanism may provide us with a
new alternative for a smooth creation of the universe, other than
the Hawking no boundary condition \cite{HH} or the Gott's
noncausal self-creating condition \cite{Gott}.

Before closing up, a quite interesting point is worth mentioning.
The checked possibility of establishing causality-violating links
between the region inside the two big rips and the regions outside
that interval for brane solutions (2.12) makes it unavoidable that
the space-time to be considered extends from $-\infty$ to
$+\infty$ without passing through the big rip singularities.
However, for this to be possible it is necessary that the scale
factor be well-defined along this infinite interval, a case which
can only be satisfied if the parameter $\epsilon$ (or $|\epsilon|$
in case of the solution containing two symmetric big bangs)
entering the equation of state is discretized so that \cite{Gonz3}
\begin{equation}
\epsilon=\frac{1}{12(n+1)}, \;\;\; n=0,1,2,3,…,\infty .
\end{equation}
Even though we do not quite understand yet the deep physical
meaning of this requirement one can still say that: (i) it leads
to a preliminary "quantization" of both the involved dynamical
quantities such as the energy density, pressure, potential energy
and scalar field, and the space-time quantities such as the
occurrence time for big bang and crunch, big rip, big trip and big
hole, and (ii) it makes any future or past event horizon to
disappear, so allowing for any amount of information to be
transferred during the big trip process and the formulation of
fundamental theories based on the definition of an S-matrix, such
as string or M theories, to be mathematically consistent. It is
tempting to speculate nevertheless that the discretization of the
equation of state parameter could be regarded to be at
qualitatively the same footing with respect to a proper quantum
theory of the universe as the original Borh theory did in relation
with the final probabilistic description of the quantum mechanics
of the hydrogen atom.

Although the present work is of a rather speculative nature, all the
mathematics and physics involved are rigorously performed and
displayed to yield clear results albeit at least some of the
interpretations that follow them turn out to be debatable. The
authors are therefore suspicious that the conclusions derived in the
present paper might play some roles related to subjects such as the
final quantum description of the universe and its future evolution.

\acknowledgements

\noindent This work was supported by MEC under Research Project
No. FIS2005-01181. The author benefited from discussions with C.
Sig\"{u}enza, A. Rozas, S. Robles,  J.A. Jim\'{e}nez Madrid, S.D.
Vereshchagin and A.V. Astashenok.

\end{document}